\documentclass[dvips]{arxstspdf}
\usepackage{mathrsfs}
\usepackage{graphicx}
\usepackage{flushend}
\usepackage{stfloats}

\volume{23}
\issue{3}
\pubyear{2008}
\firstpage{420}
\lastpage{438}
\doi{10.1214/07-STS248}

\setattribute{abstract}   {width}  {33pc}
\setattribute{keyword}    {width}  {33pc}

\begin{document}
\begin{frontmatter}

\title{A Conversation with Myles Hollander}
\runtitle{A Conversation with Myles Hollander}

\begin{aug}
\author[a]{\fnms{Francisco J.} \snm{Samaniego}\corref{}\ead[label=e1]{fjsamaniego@ucdavis.edu}}
\runauthor{F. J. Samaniego}

\affiliation{}

\address[a]{F. J. Samaniego is Professor, Department of Statistics,
University of California, Davis, California 95616, USA, \printead{e1}.}

\end{aug}

%
\begin{abstract}
Myles Hollander was born in Brooklyn, New York, on March 21, 1941. He
graduated from Carnegie Mellon University in 1961 with a B.S. in
mathematics. In the fall of 1961, he entered the Department of
Statistics, Stanford University, earning his M.S. in statistics in 1962
and his Ph.D. in statistics in 1965. He joined the Department of
Statistics, Florida State University in 1965 and retired on May 31,
2007, after 42 years of service. He was department chair for nine years
1978--1981, 1999--2005. He was named Professor Emeritus at Florida State
upon retirement in 2007.

Hollander served as Editor of the \textit{Journal of the
American Statistical Association, Theory and Methods}, 1994--1996, and
was an Associate Editor for that journal from 1985 until he became
\textit
{Theory and Methods} Editor-Elect in 1993. He also served on the
editorial boards of the \textit{Journal of Nonparametric Statistics}
(1993--1997; 2003--2005) and \textit{Lifetime Data Analysis} (1994--2007).

Hollander has published over 100 papers on nonparametric
statistics, survival analysis, reliability theory, biostatistics,
probability theory, decision theory, Bayesian statistics and
multivariate analysis. He is grateful for the generous research support
he has received throughout his career, most notably from the Office of
Naval Research, the U.S. Air Force Office of Scientific Research, and
the National Institutes of Health.

Myles Hollander has received numerous recognitions for his
contributions to the profession. He was elected Fellow of the American
Statistical Association (1972) and the Institute of Mathematical
Statistics (1973), and became an elected member of the International
Statistical Institute (1977). At Florida State University he was named
Distinguished Researcher
Professor (1996), he received the Professorial Excellence Award (1997),
and in 1998 he was named the Robert O. Lawton Distinguished Professor,
an award made to only one faculty member per year and the University's
highest faculty honor.

Myles Hollander was the Ralph A. Bradley Lecturer at the
University of Georgia in 1999, and in 2003 he received the Gottfried E.
Noether Senior Scholar Award in Nonparametric Statistics from the
American Statistical Association.
He was the Buckingham Scholar-in-Residence at Miami
University, Oxford, Ohio in September, 1985, and had sabbatical visits
at Stanford University (1972--1973; 1981--1982), the University of
Washington (1989--1990) and the University of California at Davis
(Spring, 2006).
The following conversation took place in Myles Hollander's
office at the Department of Statistics, Florida State University,
Tallahassee, on April~19,~2007.

\end{abstract}

\begin{keyword}
\kwd{Nonparametrics}
\kwd{Bayesian methods}
\kwd{Dirichlet process}
\kwd{biostatistics}
\kwd{ranking methods}
\kwd{reliability theory}
\kwd{stochastic comparisons}
\kwd{system signatures}
\kwd{writing}
\kwd{editing}
\kwd{administration}.
\end{keyword}

\end{frontmatter}

\section*{Why Statistics?}

\textbf{Samaniego:} It's a real pleasure to be back at Florida State,
Myles. I spent my first postdoctoral year in the Statistics Department
here, and I have many fond memories. Though we've been friends for over
35 years, there are many details of your life and career that I'm
looking forward to hearing more about. Let's start somewhere near the
beginning. I know that you began your college career at Carnegie Mellon
as an engineering major. Can you tell me how you got interested in
Statistics?

\textbf{Hollander:} I came to Carnegie Mellon, it was\break Carnegie Tech
when I entered in 1957, with the aim of becoming a metallurgical
engineer, but all the engineering students took more or less the same
curriculum, including calculus, chemistry, English, history of western
civilization. As the year progressed I found I liked math and chemistry
the best so near the end of the year, I went to see the heads of
metallurgy and math. The metallurgy chair was informative but laid back
and said it was my decision. The math chair, David Moscovitz, was much
more enthusiastic. He said, ``Hollander, we want you.'' Well, I was
only 17, impressionable, and I liked being wanted so I became a math
major. I didn't encounter a formal course in statistics until my junior
year. That year, Morrie DeGroot (who had come to Carnegie the same
year I did---1957---he with a Ph.D. from the University of Chicago)
taught a course that I really enjoyed. It was based on Mood's
``Introduction to the Theory of Statistics.'' DeGroot wrote some
encouraging comments on a couple of my exams and I began thinking I
might become a statistician. Then in my senior year, I took two more
excellent statistics courses from Ed
Olds. Olds at that point was a senior faculty member who had actually
done some work on rank correlation but was, I think, more known for his
consulting with nearby industry, Westinghouse, U.S. Steel and others.
In the
afternoon he taught a statistical theory course from Cram\'er's
``Mathematical Methods in Statistics.'' In the evening he taught a
course on quality control. I liked the juxtaposition of beautiful
theory that could also be useful in an
important applied context. I would say those three courses, those two
teachers, sealed the deal for me. Carnegie wanted me to stay on and do
my Ph.D. there in the math department but the lure of California, Palo
Alto, Stanford's
statistics department, was too great, so I headed west.

\textbf{Samaniego:} Let me ask a quick question about the books you
mentioned. Cram\'er is even today thought of as a very high-level book
mathematically. It's surprising that it was used in an undergraduate
course.

\textbf{Hollander:} In retrospect it is surprising but Olds taught a
beautiful course and it helped me later on in my studies. I still have
the book in my library and I look at it from time to time.

\textbf{Samaniego:} I see it and it's clearly well worn.

\textbf{Samaniego:} You were attracted to math and science in your
early years. Was that your main focus in high school?

\textbf{Hollander:} I was on an academic track in high school and
studied mostly math and science. I attended an excellent public high
school, Erasmus Hall, in the heart of the Flatbush Avenue section of
Brooklyn. It was a three-block walk from my apartment house. Naturally,
I also took other types of courses, English, social studies, history,
mechanical drawing, and
Spanish. Math was my best subject and that seemed fortunate for a kid
who wanted to be an engineer.

\textbf{Samaniego:} How did a kid from Brooklyn end up choosing to go
to a private college in Pittsburgh? I suppose that once the Dodgers
left town, you felt free to leave, too.

\textbf{Hollander:} I could have stayed in Brooklyn and gone to
Brooklyn College, thereby saving a lot of money. I could have stayed in
New York State and gone to Rensselaer Polytechnic Institute, where
several of my close friends chose to go. I wanted something different,
and Pittsburgh, despite its reputation then as a smoggy city, due to
the steel industry, appealed to me. That the Dodgers were leaving
Brooklyn the same time I was (1957 was their last season in Ebbets
Field and also my senior year of high school) didn't affect my
thinking. I did get to see them play a few times at Forbes Field in
Pittsburgh during my years at Carnegie. Forbes Field was actually a
short walk from Carnegie and you could enter the ball
game for free after the seventh inning.

\textbf{Samaniego:} Tell me about your parents and their influence on
your academic development.

\textbf{Hollander:} My mom and dad were committed to education, wanted
me to go to college, and worked hard to make it happen. My dad had one
year of college. He was at Brooklyn Polytechnic Institute in the
1927--1928 academic year majoring in civil engineering. Then the
following year the Depression hit and my father, as the oldest of three
siblings, went to work to help support his family. He never got back to
college. My dad went on to open a sequence of haberdashery stores,
mostly selling pants and shirts, in the boroughs of Manhattan, Queens
and Brooklyn. My mother did not have college training but worked as a
bookkeeper, mostly for a firm that managed parking lots throughout the
city. They both left early in the morning and came back at dinner time.
I was a latch-key kid before the term became popular.

I lived on the first floor of an apartment house on Linden
Boulevard, directly across the street from a branch of the Brooklyn
Public Library. The library was a good place to study and in my senior
year I would thumb through books on engineering. Civil, mechanical,
electrical, aeronautical were the popular areas but metallurgy appealed
to me: the chemistry labs, blast
furnaces, protective masks, etc. I looked for schools that offered it
and I also thought that by applying to a less popular field, I would
increase my chances of being accepted, and getting a scholarship.

\textbf{Samaniego:} I know you had scholarship support from the Ladish
Forging Company while at Carnegie Mellon, and also worked for them in
the summers. What was the work like? Did it play a role in your
decision to go to graduate school?

\textbf{Hollander:} When I switched from metallurgy to math at the end
of my freshman year, I contacted the Ladish Forging Company. They said
that was fine, they would still support me, which I obviously
appreciated. Then in the summer of my junior and senior years I worked
for them in Cudahy, Wisconsin. I estimated the costs of drop forgings
using the costs of
materials, the geometrical shapes of the parts, labor costs. I did some
of that each summer and also wrote some programs in Basic for the IBM
1401. My supervisor told me on the parts I estimated for which the
company was low bidder, the company lost money. I was biased low. But
he said it was fine because the workers needed the work. Ladish
actually wanted me to work for them after graduation but I wanted to
study statistics and my heart was set on Stanford. Ladish wasn't my
last position in the private sector. In the summers of 1962--1963, after
my first and second years of grad school, I worked for the Sylvania
Reconnaissance Laboratories in Mountain View. There I did get to use
some of the material I was learning at Stanford, particularly Markov
chains and stochastic processes. In the summer of my junior year, I had
an internship at the Presbyterian Medical Center in San Francisco.
Gerry Chase and I rode the Southern Pacific Railroad from Palo Alto to
San Francisco two or three times a week and worked on medical data.
Nevertheless, even though I liked these summer jobs, as my years in
graduate school increased my inclination to join the private sector
decreased.

\begin{figure}

\includegraphics{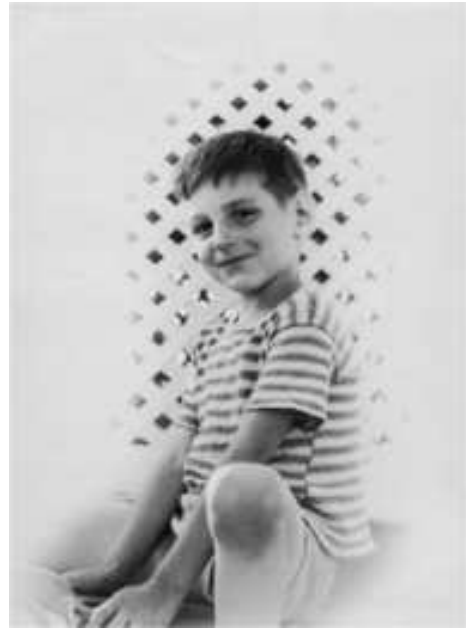}

\caption{Myles Hollander at age 6, Brooklyn, New York, 1947.}
\end{figure}

\begin{figure}[b]

\includegraphics{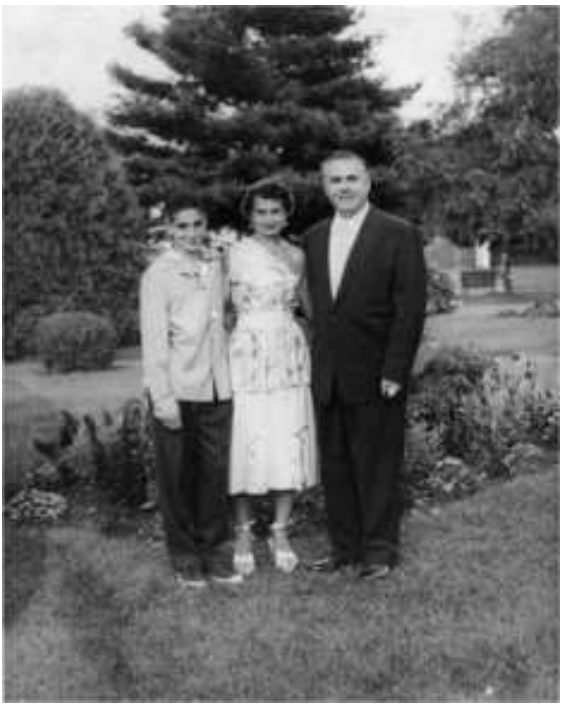}

\caption{Myles Hollander, with his parents Ruth and Joseph
Hollander, Catskill Mountains, New York, 1954.}
\end{figure}

\begin{figure}

\includegraphics{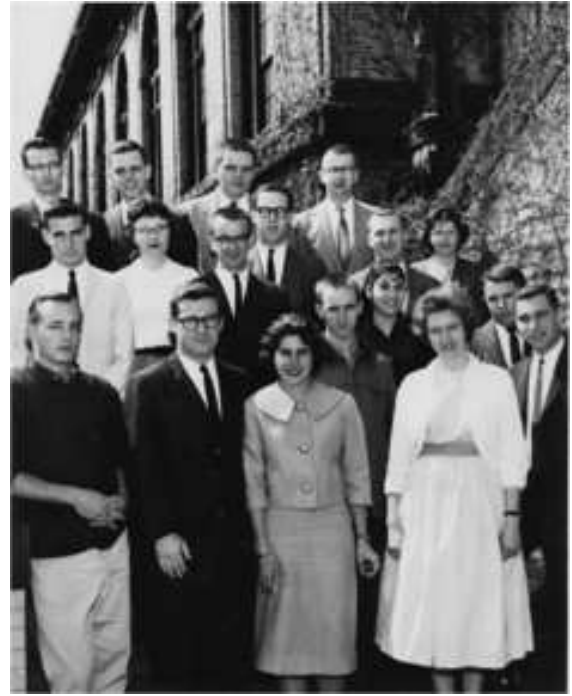}

\caption{Myles Hollander with the graduating class of the
Mathematics Department, Carnegie Institute of Technology, 1961.}
\end{figure}

\section*{Graduate School at Stanford}

\textbf{Samaniego:} Your graduate study at Stanford heavily impacted
your career choices and the statistical directions you have taken. Tell
me about your cohort of students at Stanford.

\textbf{Hollander:} It was a terrifically talented cohort. Brad Efron,
Howie Taylor, Joe Eaton, Carl Morris, Grace Wahba, Barry Arnold, Jim
Press, Paul Holland, Jean Donio, Galen Shorack, Gerry Chase, and many
more. I should really name them all. We were all excited about the
material. We wanted to learn what our professors taught and we wanted
to learn how to do it
ourselves. We were very cooperative and friendly among ourselves. I
have many memories, Howie Taylor working on (and talking about) a
probability problem at the blackboard in our office in Cedar Hall, Carl
Morris and I talking about Pitman efficiency at a blackboard in an
empty classroom in Sequoia Hall and Carl shedding light on what was
going on, Barry Arnold
and I discussing a mathematical statistics problem in Cedar---many, many
such instances. Brad Efron was a senior student to our group who
interacted with us and helped us in many ways, including discussing
geometrical interpretations of theorems. We typically took the
qualifying exams in the middle of our third year. To help us prepare,
we would each choose a topic
and write a 10--12-page focused summary with solutions to problems,
theorems, key ideas. I did one on nonparametrics, Howie Taylor did one
on advanced probability, and so forth. We put the summaries together,
made copies and passed them amongst ourselves. When we took our orals
we were pumped, prepared, and, to the extent that one can be for such a
momentous test, we were confident. Also, of course, we were nervous. My
exam committee was Lincoln Moses, Rupert Miller, Charles Stein and Gerry
Lieberman and I see them sitting there today just as I am looking at
you and I remember most of the questions to this day.

\textbf{Samaniego:} Give me an example of a question that was asked.

\textbf{Hollander:} Well, Lincoln Moses asked about nonparametric tests
for dispersion and I decided to mention one of his rank tests. Then
Gerry Lieberman turned to Lincoln and said in mock surprise, ``Lincoln,
you have a test?'' They were close friends so Gerry could tease him in
this way but Lincoln wasn't particularly happy about my answer and then
he threw a
tough question at me about the asymptotic distribution of the
Kolmogorov--Smirnov statistic. Charles Stein asked me about decision
theory and I was ready for that. I went to the blackboard and outlined
the framework of a decision theory problem just like he did at the
beginning of many of his lectures.

\textbf{Samaniego:} He didn't ask any testy inadmissibility questions,
did he?

\textbf{Hollander:} I had covered the blackboard and used a lot of
time but he did ask about the relationship between admissibility and
invariance. It had been covered in his course so I was ready for it.

\textbf{Samaniego:} Which faculty members at Stanford had the greatest
influence on you, personally and professionally?

\textbf{Hollander:} Lincoln Ellsworth Moses had the greatest
influence. I was lucky at the start because my first TA assignment in
fall quarter, 1961, was to be a grader in the elementary decision
theory course he was teaching out of Chernoff and Moses. He gave the
main lectures and five or six TAs graded papers and met with sections to go
over homework. I got to know Lincoln through this activity and he also
encouraged me to attend the biostatistics seminar that he and Rupert
Miller were giving in the medical school. I would also be invited to
his home in Los Trancos Woods and got to know his wife Jean and their
children. I was close to him throughout and after he married Mary Lou
Coale, Glee and I remained very close with them. Beginning in the fall
of 1963, Lincoln taught a two-quarter course on nonparametric
statistics. It was a beautiful contemporary sequence and there was lots
of nonparametric research in that period, particularly by Erich Lehmann
and Joe Hodges at Berkeley, Lincoln, Rupert Miller, Vernon Johns at
Stanford. Lincoln named me the TA for that course even though I was
taking it at the same time. There I was, grading the papers of my
really talented fellow students, like Joe Eaton---and so I had to be
good. I was determined to excel, to be one of the best if not the best
in the class. Later,
motivated by this course, I wrote a thesis on nonparametrics under
Lincoln's direction.

Lincoln became my role model, the statistician I most admired
and tried to emulate. He showed me how to be a professional, the joy of
statistics, and the great pleasure of being a university professor. In
my career I have tried to do for my students what Lincoln did for me.

\textbf{Samaniego:} What you say about Lincoln Moses rings very true.
From my own few interactions with him, and from things I've heard about
him over the years, he was both a fine teacher and scholar and a true
gentleman. Tell me about your interactions with other Stanford
faculty.

\textbf{Hollander:} I was also strongly influenced by other professors
from whom I took courses. Rupert Miller via the biostatistics seminar,
Ingram Olkin through the problems seminar he co-taught with Shanti
Gupta, who was visiting in 1961 (they started out assigning problems in
Cram\'er's book and that was a break for me as a beginning student
because I had seen most of the problems at Carnegie). Ingram also
taught multivariate analysis which I also took. I took Charles Stein's
decision theory sequence and Manny Parzen's time series sequence.
Kai-Lai Chung taught the advanced probability sequence. They were all
dedicated to their subjects, made them come alive, each had his own
style, and each was at the top of his game. Then on two sabbaticals at
Stanford, working in the medical center, I became friendly with Bill
and Jan Brown and reinforced my friendship with Rupert and Barbara
Miller. Bill and Jan became the godparents to our children. One special
bond that existed between Rupert and Barbara and Glee and me: Jennifer
Ann Miller and Layne Q Hollander were delivered the same day, October
29, 1964, at Stanford Hospital, and
Glee and Barbara shared the same hospital room for three or four days. Over
the years, I've grown closer to Ingram through the various
international conferences on reliability that you and I have attended
and to Manny through his work with the nonparametrics section of ASA.

\textbf{Samaniego:} All of the people that you've mentioned have
written very good books in probability or statistics. I'm wondering,
since you've co-authored three books yourself, whether these people and
the way they wrote influenced you?

\textbf{Hollander:} I did put a high premium on clear writing in the
three books I've co-authored. I think the person who influenced me the
most in that regard was Frank Proschan, who insisted on clear writing.
When I took the course on stochastic processes, it was based on Manny's
notes (his book was not yet out) and it was taught by Don Gaver. When I
took Ingram's multivariate analysis course, he used his notes and Ted
Anderson's book. I used Rupert's book on multiple comparisons for
research, but I didn't take that subject as a course. Kai-Lai used the
notes that would become his beautiful
book on advanced probability. Certainly Manny, Ingram, Rupert and
Kai-Lai wrote in clear, captivating ways.

\textbf{Samaniego:} You met your wife Glee at Stanford and the two of
you were married in the Memorial Church on the Stanford campus. Many of
your friends feel that your bringing Glee into the extended Statistics
community is your greatest contribution to the field! Tell me how you
met Glee and how you managed to persuade her to marry you. (Laughs)

\textbf{Hollander:} I was sitting in my office on the second floor of
Ventura Hall at Stanford in October, 1961. It's a spacious office and
even though it had four desks, only two students would come regularly,
Jon Kettenring and me. (A year later Pat Suppes would take over that
office.) I was working on a hard problem and I paused to look out the
window. I saw a young girl walking briskly, determined, in high heels,
with blond hair, bouncing along with remarkable energy (past Ventura,
maybe to the Computer Center). A California girl! Clearly I could never
even approach a person like that. She passed out of my view and I went
back to my homework, probably a waiting time problem in stochastic
processes. The expected waiting time for me to approach the girl I had
just seen was no doubt infinite.

Eight months later, in June, 1962, my friend Heinz, an
engineering student from Germany, and I decided to go on a double date.
We decided to meet on a Friday night at El Rancho, a restaurant on El
Camino Real, in Palo Alto. In addition to dinner, El Rancho also had a
dance floor and a lively band. When I arrived I realized that Heinz's
date was unmistakably the girl I had seen when gazing from my Ventura
Hall office in the fall---Glee.

The evening was going well and I was totally enthralled by
Glee, her brightness, her wit, her energy, her enthusiasm, her bounce.
After about an hour the band played ``It's Cherry Pink and Apple
Blossom White''---a cha cha. I asked my date to dance but she said she
didn't cha cha. I mustered the courage to ask Glee. She said, ``I'll
try.'' Of course she was and still is a great dancer and I was on cloud
nine. I thought I'd made a good impression. A week later I called her
on the phone and said, ``Hi Glee, it's Myles Hollander.'' She said,
``WHO?'' Obviously I did not impress her as much as she had impressed
me. Clearly I needed to go into high gear. I took her sailing on San
Francisco Bay. I took her horseback riding in the foothills behind
Stanford. I took her skiing at Heavenly Valley. Eventually my
persistence triumphed. We hit it off over a period of about a year, and
got married at Stanford Memorial Church on the Stanford campus in
August, 1963. We went on to have two fine sons, Layne Q and Bart Q,
who, with their wives, Tracy and Catherine, also gave us five wonderful
grandchildren---Taylor, Connor, Andrew, Robert and Caroline. Glee
earned her Ph.D. at FSU in an excellent clinical psychology program
and worked in private practice, and also at Florida State Hospital in
Chattahoochee. I like to say it all started with the cha cha and we're
still dancing after all these years!

\textbf{Samaniego:} On the statistical front, you published a major
portion of your thesis in a pair of \textit{Annals} papers. What was the
main focus of this work?

\textbf{Hollander:} My thesis was devoted to rank tests for ordered
alternatives in the two-way layout. Lincoln Moses, in his nonparametric
sequence in the third year of my graduate work, had covered ordered
alternatives in the one-way layout and that suggested to me some ideas
for randomized blocks. I proposed a test based on a sum of overlapping
signed rank statistics that is not strictly distribution-free but can
be made asymptotically distribution-free. Kjell Doksum at Berkeley was
also working on closely related problems at the same time and in the
end our two papers were published adjacently in the 1967 \textit{Annals}
(Doksum, \citeyear{d67}; Hollander, \citeyear{h67}). In my thesis I also pointed out a
certain multiple comparison procedure, thought by Peter Nemenyi
(Nemenyi, \citeyear{n63}) to be distribution-free, was not, but could
be made asymptotically distribution-free. I published the
asymptotically distribution-free multiple comparison procedure in the
1966 \textit{Annals} (Hollander, \citeyear{h66}).

\begin{figure}

\includegraphics{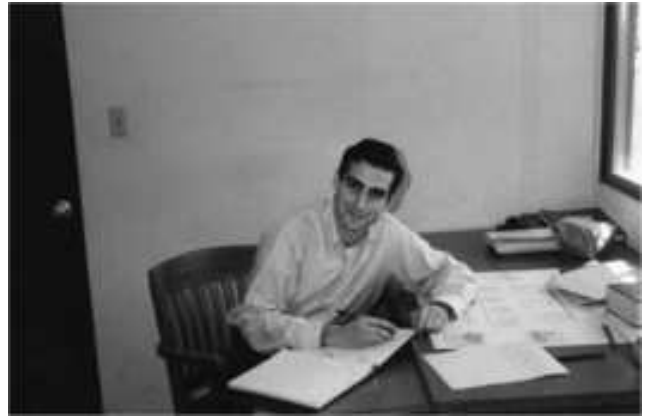}

\caption{Myles Hollander as a graduate student at Stanford, 1963.}
\end{figure}

\begin{figure}[b]

\includegraphics{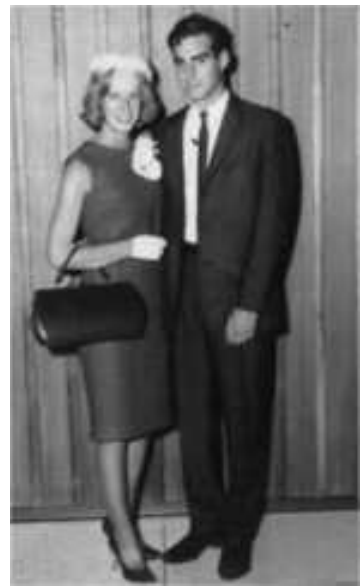}

\caption{Myles and Glee Hollander, starting life together after
being married at Stanford Memorial Church, August 17, 1963.}
\end{figure}

\section*{A Career at Florida State University}

\textbf{Samaniego:} You've written quite a few papers on classical
nonparametric testing problems. Give us an idea of the range of
problems you have worked on in this area.

\textbf{Hollander:} In my early years at FSU I wrote nonparametric
papers on bivariate symmetry, regression, uncorrelated nonparametric
statistics, and did a little more on ordered alternatives. I also
worked with my first Ph.D. student, Ron Randles, on a paper that was
decision-theoretic rather than nonparametric. We developed $T$-minimax
procedures for selection
procedures and it was published in the 1971 \textit{Annals} (Randles and
Hollander, \citeyear{rh71}). Ron took my class in nonparametrics and even though
his thesis was not nonparametric in character, he did excellent work,
went on to be a leader in nonparametrics and set a very high bar for my
subsequent Ph.D. students.

Thus in the beginning I was working on my own and with
students. That was the way the senior leaders in the department, Ralph
Bradley and Richard Savage, wanted it. Work on your own, prove your
mettle, and move away from your thesis topic. Later on, when I began to
collaborate with Frank Proschan and Jayaram Sethuraman, two great
statisticians, my scope of topics vastly increased and my research got
better! Whenever I received an offer or feeler from another place, I
had to ponder whether I could find and establish working relationships
with such superb collaborators at the next stop. I always\break doubted it.

\textbf{Samaniego:} Your research over the years has been distinctly
nonparametric, including, of course, interesting and important
contributions to Bayesian nonparametrics. You and your doctoral
student,
Ramesh Korwar, were the first to develop inference procedures for the
hyperparameter of Ferguson's Dirichlet process, establishing the
foundations for an empirical
Bayes treatment of nonparametric estimation. I see that it's an
interest you've sustained up to the present time. How did you get
interested in this latter problem area?

\textbf{Hollander:} My interest in the Dirichlet process arose from
Tom Ferguson's seminal paper (Ferguson, \citeyear{f73}). That was the principal
motivation. I had obtained a preprint before its publication. I had
read some earlier papers at Stanford on Bayesian nonparametrics but
Ferguson's paper was the most tractable, the most promising. I can't
remember the exact timing but I went to a Bayesian nonparametric
conference at Ohio State where Tom was the principal speaker. He was
also aware of some of the results by Ramesh Korwar and me and mentioned
them in his lectures. His wonderful lectures got me further fired up
and I went on to do more Bayesian nonparametrics with Ramesh, and then
later with two more of my Ph.D. students, Greg Campbell and Bob Hannum,
and more recently with Sethu (Campbell and Hollander, \citeyear{ch78}; Hannum,
Hollander and Langberg, \citeyear{hhl81}; Hannum and Hollander, \citeyear{hh83}; Sethuraman
and Hollander, \citeyear{sh07}).

\textbf{Samaniego:} Which ideas or results in your Bayesian
nonparametric papers seem to have had the most impact?

\textbf{Hollander:} Ramesh Korwar and I had several interesting
results in our 1973 paper in the \textit{Annals of Probability} (Korwar and
Hollander, \citeyear{kh73}). We showed that when the parameter $\alpha$ of the
Dirichlet process is nonatomic and $\sigma$-additive, $\alpha(\mathscr
{X})$ can be estimated from a sample from the process. The estimator we
devised is $D/\log(n)$, where $D$ is the number of distinct
observations in the sample. We proved that estimator converges almost
surely to $\alpha(\mathscr{X})$ where $\alpha$ is a finite nonnull
measure on a space $\mathscr{X}$ that comes equipped with a $\sigma
$-field of subsets. We also showed in the nonatomic and $\sigma
$-additive case, that given $D$, the $D$ distinct sample values are
i.i.d.
with distribution $\alpha(\cdot)/\alpha(\mathscr{X})$. This result has
been used by others. For example, in an \textit{Annals} paper Doksum and
Lo (Doksum and Lo, \citeyear{dl90}) considered Bayes procedures when $F$ is chosen
by a Dirichlet prior and used the result to study consistency
properties of posterior distributions.

Another result that Ramesh and I had in that 1973 paper gave
the joint distribution of the indicators that tell if the $i$th
observation is distinct from the previous $i-1$. The indicators are
independent, but not identically distributed, Bernoulli random
variables. Diaconis and Freedman (Diaconis and Freedman, \citeyear{df86}) used
this result in their study of inconsistent Bayes estimators of
location. In our 1976 \textit{Annals} paper (Korwar and Hollander, \citeyear{kh76})
Ramesh and I used the Dirichlet process to define a sequence of
empirical Bayes estimators of a distribution function. One interesting
consequence of that paper was a result reminiscent of the famous
James--Stein result on the inadmissibility of multivariate $\bar{X}$
when the dimension is $\geq3$. Ramesh and I showed that if there are
at least three distribution functions to be estimated, one could do better
than estimating each distribution by its sample distribution.

In a 1981 \textit{Annals of Probability} paper (Hannum,
Hollander and Langberg, \citeyear{hhl81}) Bob Hannum, Naftali Langberg, and I
studied the distribution of a random functional $\int Z \;dP$ of a
Dirichlet process. We related the cumulative distribution of that
functional evaluated at $x$, say, to the distributions of random
variables $T^x$ and we obtained the characteristic function of $T^x$.

It has been surprising and gratifying to see some recent uses
of this result. For example, it is used (Cifarelli and Melilli, \citeyear{cm00})
to study the distribution of the variance functional. The result is
also used (Regazzini, Guglielmi
and Di Nunno, \citeyear{rgdn02}) to study the probability distribution of the
variance of a Dirichlet measure and the probability distribution of the
mean of a Dirichlet measure. Thus the result is getting a little play
in the Italian school.

\textbf{Samaniego:} You've been at Florida State for 42 years! I'd
like to ask you about your extensive and fruitful collaborations with
some of your colleagues here. Tell me about your first joint paper with
Frank Proschan. It was, I believe, one of the first papers in which
tests were developed to detect particular nonparametric (NBU)
alternatives to the exponential distribution.

\textbf{Hollander:} Frank came here in 1971 from the Boeing Research
Labs. He was very open, very dedicated to his research. Our offices
were close and we became friends. One day he walked into my office and
said, ``Let's write a paper.'' I said, ``Great.'' I was excited he
asked. His main area was reliability and mine was nonparametric
statistics, so we aimed to work in the intersection, namely
nonparametric methods in reliability. The first paper we wrote covered
our NBU (new better than used) test (Hollander and Proschan, \citeyear{hp72}). The
test is based on a $U$-statistic, partially reminiscent of
the Wilcoxon--Mann--Whitney statistic. We enjoyed working on it and there
was a mild surprise. In calculating the probability that the statistic
assumes its maximum value, the Fibonacci sequence pops up. The sequence
had not arisen in Frank's longer research experience, nor in my shorter
one. It is nice to have a mild
connection with a famous pre-Renaissance mathematician. I believe the
paper stimulated more research in testing and estimation for the
various nonparametric classes arising naturally in reliability,
including more research avenues for us.

\textbf{Samaniego:} In a subsequent paper, you and Frank discovered an
interesting new context in which the total-time-on-test statistic
arose. I'm sure that was a pleasant surprise.

\textbf{Hollander:} Frank and I wrote a testing paper on mean residual
life (Hollander and Proschan, \citeyear{hp75}) that was published in \textit{Biometrika}.
We considered the decreasing mean residual life class, the new better
than used in expectation class, and their duals. We defined measures of
DMRLness and NBUEness based on $F$, plugged in the empirical for $F$,
and used those plug-in statistics as test statistics, standardized to
make them scale-invariant. In the NBUE case, we obtained the
total-time-on-test statistic. Up to that time it had been viewed as a
test of exponentiality versus IFR or IFRA
alternatives. We showed its consistency class contained the larger set
of NBUE distributions, thus broadening its interpretation and
applicability.

Large nonparametric classes of life distributions captured
our attention for awhile. For example, we co-directed our student Frank
Guess on a project where we defined new classes relating to a trend
change in mean residual life. In our 1986 \textit{Annals} paper (Guess,
Hollander and Proschan, \citeyear{ghp86}) we considered the case where the change
point is known. Later (Kochar, Loader and Hawkins, \citeyear{klh92}) procedures
were given for the situation where the change point is unknown.

\textbf{Samaniego:} On a personal level, what was it like to
collaborate with Frank Proschan? Give us a feeling for his sense of
humor, his work ethic, the ``Reliability Club'' and his overall
influence on you.

\textbf{Hollander:} Frank, as you know, had a deadpan sense of humor.
He would often remind me of the comedian Fred Allen who was very funny
but never cracked a smile, never laughed at his own jokes. When he gave
a lecture Frank would adroitly use transparencies, and there was always
a parallel processing taking place, the material in the lecture, and humorous
asides. He was dedicated to his research. He would come to the office
very early, work for a few hours, go to the university pool for a swim,
go home for lunch, then come back and work again. We both would come in
on Saturday mornings, talk about what we wanted to show, go back to our
offices, try to get a result, write up the progress, then put a copy in
the other
person's mailbox. This went back and forth. Some mornings we would come
in and do this without talking face to face. The results would
accumulate, and then we would have a paper. Later, Frank started the
Reliability Club which met on Saturday mornings to present and discuss
topics on reliability. Many students, several faculty and visitors
would attend, and it would lead to dissertations, joint work, research
grants, more papers. I had the habit of working some weekends
(including some Sunday nights; Glee and I lived very close to campus
then, about an 8-minute drive) before Frank arrived but Frank
solidified it and showed me I was not crazy doing it (or else we were
both crazy). Without trying or fully realizing it,
Frank's style and work ethic became a part of mine.

\textbf{Samaniego:} You've nicely integrated the parallel processing
of material and humor into your own presentation strategy.

\textbf{Hollander:} Frank, I've always tried to be funny. It's both a
strength and a weakness. I like to make people laugh but every once in
a while it's not the time to be funny. Over the years I've become
better at resisting the temptation to
try to say something funny. But I still like to make witty remarks. I
like to present to people the notion that statisticians have pizazz.

\textbf{Samaniego:} I've found myself that in teaching our subject, a
little bit of well-timed humor---not the stand-up comedy type but the
things that actually have something to do with the material we are
talking about---helps people stay aboard; most people listen and enjoy
it.

\textbf{Hollander:} At this point your advice on how to teach gets
much higher marks than mine because you have just won an outstanding
award at the University of California, Davis. I won't even mention the
figure here; otherwise people will come by your house at night and
break in.

\textbf{Samaniego:} Well, Myles, I've always enjoyed your presentation
style and have probably stolen more than I care to acknowledge from the
talks I've heard you give.

\textbf{Samaniego:} You've written a good many papers with Jayaram
Sethuraman. What would you consider to be the highlights of that work?

\textbf{Hollander:} Even before Sethu and I worked together, I would
go to his office for consulting. He is a brilliant statistician and he
can often point you in a direction that will help, or lead you to a
breakthrough when you are
stuck. His entire career he has been doing that for all those wise
enough to seek his assistance.

My first paper with Sethu is also joint with Frank. It is the
DT (decreasing in transposition) paper (Hollander, Proschan and
Sethuraman, \citeyear{hps77}) and is certainly a highlight. It is a paper on
stochastic comparisons which yields many monotonicity results. Among
the applications in that paper were power inequalities for many rank
tests. Al Marshall and Ingram Olkin later changed the DT term to AI
(arrangements increasing). In a later paper (Hollander and Sethuraman,
\citeyear{hs78}), Sethu and I gave a solution to a problem posed to us by Sir
Maurice Kendall during his short visit to Tallahassee in 1976. It was
``How should one test if two groups of judges, each giving a complete
ranking to a set of $k$ objects, agree, that is, have a common
opinion?'' We proposed a conditionally distribution-free test using the
Wald--Wolfowitz statistic.

\textbf{Samaniego:} Tell me about your more recent work with Sethu.

\textbf{Hollander:} Sethu and I have, on and off, been working on
repair models in reliability for the last 15 years. Our interest was
sparked by your groundbreaking paper with Lyn Whitaker (Whitaker and
Samaniego, \citeyear{ws89}) in which you developed what is now called the
Whitaker--Samaniego estimator of the distribution $F$ of the time to
first repair in imperfect repair
models. With Brett Presnell, we considered the problem in a counting
process framework (Hollander, Presnell and Sethuraman, \citeyear{hps92}) and also
developed a simultaneous confidence band for $F$ as well as a
Wilcoxon-type two-sample test in the repair context. Many other
important parameters, such as the expected time between repairs, depend
on $F$ and the nature of the repair process, so the problem of
estimating $F$ is important.

Five years later, with Cris Dorado (Dorado, Hollander and
Sethuraman, \citeyear{dhs97}) we proposed a very general repair model that contains
most of the models in the literature. We also introduced the notion of
life supplements or boosts, so not only could the repairman move the
effective age of the system to a point better than, say, minimal repair,
he could also boost the residual life.

Recently we finished a paper on Bayesian methods for repair
models (Sethuraman and Hollander, \citeyear{sh07}). For example, if you put a
Dirichlet prior on $F$ in, say, the imperfect repair model, and take
two
observations, the posterior distribution of $F$ is no longer Dirichlet.
Thus there is, for these complicated repair processes which induce
dependencies, a need for a broader class of priors which are conjugate.
We introduced partition-based priors and showed they form a conjugate
class. Beyond repair models, we believe this new method for putting
priors on distributions has potential in many other areas.

\textbf{Samaniego:} One of my favorites among your papers is a JASA
paper you wrote with Chen and Langberg on the fixed-sample-size
properties of the Kaplan--Meier estimator. It was based on a simple but
very clever idea. Can you describe that work and how it came about?

\textbf{Hollander:} I was interested in the KME's exact bias and its
exact variance. Brad Efron (Efron, \citeyear{e67}), in his fundamental article on
the two-sample problem for censored data, had given bounds on the bias.
Proportional hazards provided a clean way to get exact results.
Earlier, Allen (Allen, \citeyear{a63}) proved that when the cumulative hazard function of
the censoring distribution is proportional to that of the survival
distribution, the variables $Z = \min(X,Y)$ and the indicator function
$I(X \leq Y)$ are independent, where $X$ is the time to failure, $Y$ is
the time to censorship. In his 1967 paper, Efron used this result for
obtaining efficiencies for his generalized Wilcoxon statistic in the
case when the censoring and survival distributions are exponential, and
he thanked Jayaram Sethuraman for bringing the result to his attention.
In the KME setting we (Chen, Hollander and Langberg, \citeyear{chhl82}) obtained an
exact expression for moments of the KME by conditioning on $Z = (Z_1,
\ldots, Z_n)$ and using Allen's result. Getting exact results in this
setting was a natural consequence of my interest in rank order
probabilities. Erich Lehmann really planted the seed with his famous work
on the power of rank tests (Lehmann, \citeyear{l53}) where he obtained exact
powers against what are now called Lehmann alternatives. My natural
tendency is to first try hard to get exact results, then move to
asymptotics.

\textbf{Samaniego:} You've done extensive joint work with some of your
doctoral students. Perhaps your collaboration with Edsel Pe\~na is the
most varied and most productive. Tell me a little about that work.

\textbf{Hollander:} Edsel is an amazingly dynamic and energetic
researcher. He loves to do research and his enthusiasm is infectious.
He is also very talented. We have worked on a broad range of problems.
We started (Hollander and Pe\~na, \citeyear{hp88}) with obtaining exact
conditional randomization distributions for various tests used to
compare treatments in clinical trials
that use restricted treatment assignment rules, such as the biased coin
design. We have also worked on confidence bands and goodness-of-fit
tests in censored data settings. For example, in our 1992 JASA paper we
(Hollander and Pe\~na, \citeyear{hp92}) defined a\break goodness-of-fit test for
randomly censored data that reduces to Pearson's classical test when
there is no censoring. We considered the simple null
hypothesis and later Li and Doss (Li and Doss, \citeyear{ld93}) extended it to the
composite case. Thus, although not ideal, there are secondary gains in
not solving the more general problem straight out. You inspire others
and your paper gets cited.

\begin{figure}

\includegraphics{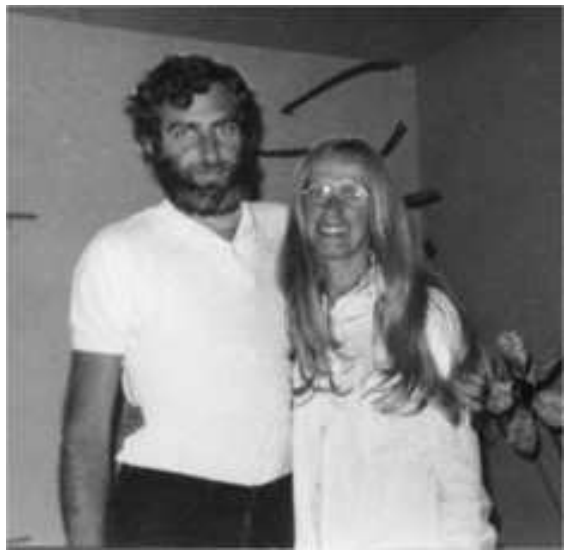}

\caption{Myles and Glee Hollander in their flower-child style on
sabbatical at Stanford, 1972.}
\end{figure}

\begin{figure}[b]

\includegraphics{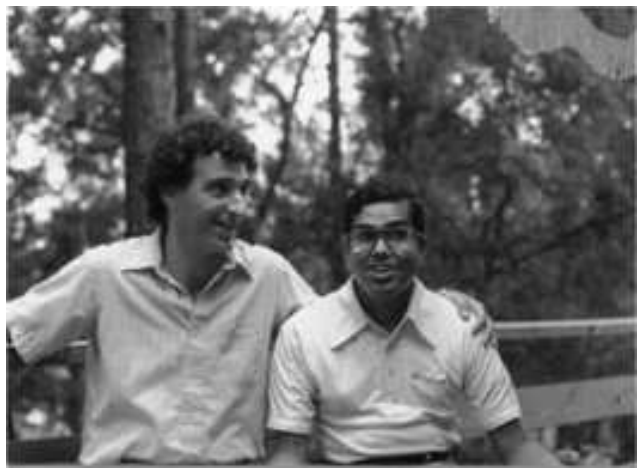}

\caption{Myles Hollander and Jayaram Sethuraman at the
Hollander's home, Tallahassee, 1978.}
\end{figure}

\begin{figure*}

\includegraphics{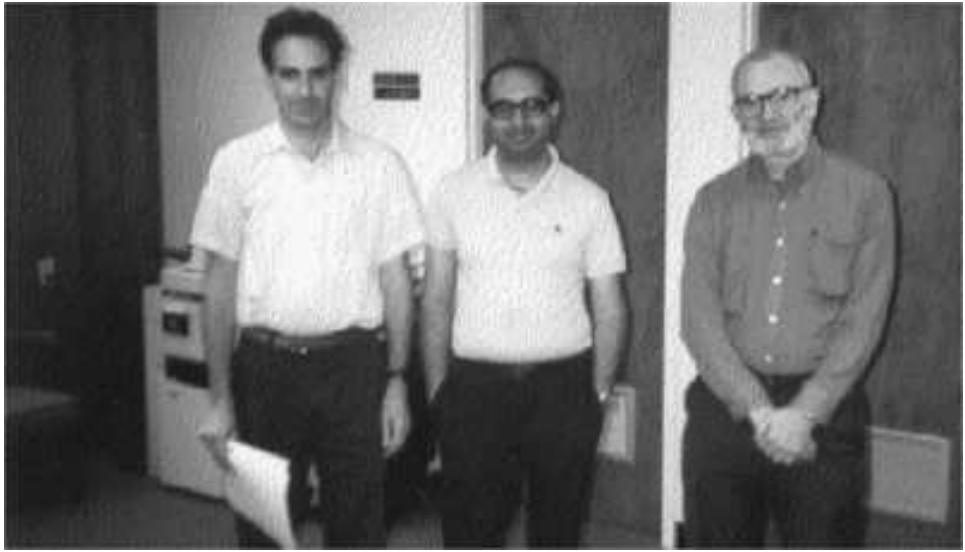}

\caption{Myles Hollander, Subhash Kochar and Frank Proschan in
the Department of Statistics, FSU, 1983.}
\end{figure*}

Edsel and I have also worked on interesting reliability
models. For example, in Hollander and Pe\~na (\citeyear{hp95}) we used a Markovian
model to describe and study system reliability for systems or patients
subject to varying stresses. As some parts fail, more stresses or loads
may be put on the still-functioning parts. We use the failure history
to incorporate the changing degrees of loads and stresses on the
components. Shortly after that (Hollander and Pe\~na, \citeyear{hp96}) we
addressed the problem about how a subsystem's performance in one
environment can be used to predict its performance in another
environment. Another idea that may attract some interest is our class
of models proposed in 2004 in the \textit{Mathematical Reliability}
volume (Pe\~na and Hollander, \citeyear{ph04}). We introduced a general class of
models for recurrent events. The class includes many models that have
been proposed in reliability and survival analysis. Our model
simultaneously incorporates
effects of interventions after each event occurrence, effects of
covariates, the impact of event recurrences on the unit, and the effect
of unobserved random effects (frailties). Edsel and his colleagues and
students have been studying asymptotic properties of the estimators and
also applying them to various data sets.

\begin{figure}[b]

\includegraphics{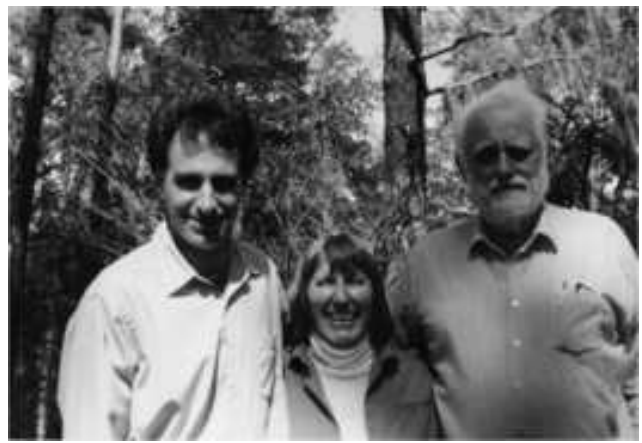}

\caption{Myles Hollander with Mary Lou and Lincoln Moses,
Tallahassee, 1985.}
\end{figure}

\section*{Other Writing}

\textbf{Samaniego:} Tell me about your three books.

\textbf{Hollander:} The nonparametric books with Doug, the first and
second editions, were very successful (Hollander and Wolfe, \citeyear{hw73}, \citeyear{hw99}). One
important feature of these books are the real examples from diverse
fields. It helped us broaden our audience beyond statisticians. Doug
and I also taught a short course for about nine years, mid'70s to 80's, at
the George Washington University Continuing Engineering Education
Center. The audience at those courses consisted mainly of people in
government and industry so again, in a way, we were bringing the
nonparametric ideas and techniques to a different audience. Wiley has
sought a third edition, but Doug and I have not yet committed to it.

Bill Brown and I began writing the medical statistics book
(Brown and Hollander, \citeyear{bh77}) in 1972 when I was on sabbatical at
Stanford. We also featured real examples and it was adopted at many
medical schools. I also used it for many years at FSU for a basic
course on statistics in the natural sciences. Wiley always wanted a
second edition, but Bill and I never got around to it. Wiley is now
going to publish the original book as a paperback in its Wiley Classics Library
series.

The book \textit{The Statistical Exorcist} with Frank Pros\-chan
(Hollander and Proschan, \citeyear{hp84}) was great fun to write. The book
consisted of vignettes that treated a variety of problems. We wrote in
a way to explain to the readers what statistics does, rather than give
a formulaic approach on how to do statistics. In fact, we didn't use
any mathematical formulas or symbols. One interesting feature is the
cartoons, about half of which were drawn by Frank and Pudge's daughter
Virginia and half drawn by Glee. Frank and I described the scenes and
supplied the captions and Ginny (Virginia) and Glee did the drawings.
We also opened the vignettes with epigraphs, relating to statistics,
from novels. Some of the epigraphs are real and some were created by
Frank and me. In an appendix we informed the reader which ones were
from our imagination. For a text, however, students found it difficult
without a few formulas upon which to hang their hats, for example, when
to multiply probabilities, when to add, and so on. Marcell Dekker also
wanted a second edition and it is not beyond the realm of possibility.

This semester I've been teaching an advanced topics course.
The material was an eclectic mixture of survival analysis and
reliability theory where I focused on some of the parallels between the
two subjects. The course title
is ``Nonparametric methods in reliability and survival analysis.''
Whenever I look at the syllabus, it occurs to me that the material
would make a good monograph. The problem is that most books on
reliability are not big sellers although some are beautiful and
informative. When I write, I do it not so much to make a few extra
dollars, but to be read and thus a potentially large audience is the
draw.

\textbf{Samaniego:} Did any specific examples in \textit{The Statistical
Exorcist} come out of your joint research with Frank?

\textbf{Hollander:} Some of the subject matter was motivated by the
joint research. For example, we had vignettes on reliability which are
unusual in an elementary book. We also had vignettes on nonparametric
statistics, so the vignettes were influenced to some extent by our
favorite subjects.

\textbf{Samaniego:} Myles, \textit{The Statistical Exorcist} is, I would
say, unique in the field as an introduction to statistical thinking.
The book is distinctive in a variety of ways including its general
content, the humor of its cartoons and epigraphs and even the titles of
some of its sections. There is one entitled, ``A Tie is Like Kissing
Your Sister.'' Tell me about that section.

\textbf{Hollander:} There was a time when college football games could
end in ties; that time has long passed and now they play extra sessions
to determine a winner. But the conventional wisdom of most coaches was
that a tie was no good. It leaves everybody frustrated and unhappy, the
players and fans on both teams. Some coach coined the phrase ``A tie is
like kissing your sister.'' Which meant, you love your sister but you
don't get much satisfaction out of kissing her. The vignette considered
an optimal strategy for near the end of the game, taking into account
the chance of making an extra point (one-point) play, the chance of
making a two-point play and the relative value of winning the game
versus the relative value of tying the game.

\section*{Administrative Work}

\textbf{Samaniego:} With all these activities, plus your teaching and
the mentoring of your graduate students, one would think that there might
have been little time for other responsibilities. But, in fact, you
served for nine years as the chair of your department. What were the
main challenges you encountered as chair, and what achievements are you
proudest of?

\textbf{Hollander:} A chair obviously has many priorities: the
faculty, the students, the staff, the administration. They are all
important and you have to serve and contribute to the well-being of
each. However, in my mind the top priority is to recruit well, get the
best people possible. Then everything desirable follows: a stronger
curriculum, research grants, better students, and so forth.

In my first term, 1978--1981, my most significant hire was Ian
McKeague who, in 1979, came from UNC, Chapel Hill. He stayed 25 years,
participated in grants, became an expert in survival analysis, and
served a three-year term as chair. We co-directed Jie Yang on a topic
on confidence bands for survival functions and have two papers that
emanated from that
work and related work on quantile functions with Gang Li (Hollander,
McKeague and Yang, \citeyear{hmy97}; Li, Hollander, McKeague and Yang, \citeyear{lhmy96}). In
my second and third terms, 1999--2005, among the tenure-earning people I
hired, Flori Bunea, from U. Washington, Eric Chicken from Purdue, Dan
McGee from University of South Carolina Medical School, and Marten Wegkamp
from Yale, seem the most likely to contribute and hopefully stay at FSU
for a long time. Each filled an important gap in our curriculum, taught
new courses, got involved with grants. I recruited Dan as a senior
biostatistician and he has been a driving force in establishing our new
M.S. and Ph.D. programs in biostatistics. He also succeeded me as
chair.

\textbf{Samaniego:} Tell me a bit about how you tried to broaden the
department's focus and reach. What are some aspects beyond recruiting?

\textbf{Hollander:} In Fall 1999 I called Ron Randles, who was
statistics chair at the University of Florida (UF) at that time, and
suggested we create an FSU-UF Biannual Statistics Colloquium Series.
Ron liked the idea and after getting approval from our faculties it
began and continues today. The idea is that it provides the opportunity
for the recent appointees of each faculty to get some outside exposure
by giving a talk in the other department. Thus in one semester UF comes
to Tallahassee and a UF person talks, and the next semester FSU goes to
Gainesville and an FSU person gives the colloquium
talk. I also hope it leads to some joint research. Some people have had
discussions, but to my knowledge it hasn't happened yet.

When I was chair, I was a mentor to all of our students, many
of whom I recruited. I tried to teach them how to become professionals.
I helped them get summer jobs and of course wrote reference letters for
them. When I was younger I played intramural basketball and softball
with some of them. I've gone to some of their weddings. Many students
still stay in close touch with me. Of course you don't have to be the
chairman to engage in these mentoring activities, but as chair one gets
many opportunities to give extra advice at, for example, orientation
and frequent student visits to the chair's
office.

Here's a chair's story that goes into the highlight category.
Ron Hobbs, an M.S. graduate of our department in 1967, and his wife
Carolyn Hobbs, who earned a B.S. in Recreation Studies from FSU in 1965,
endowed a chair in our department. It worked like this. Each year for
six consecutive years, Ron and Carolyn contributed \$100,000. Then
after six years, the state contributed \$400,000. Then the university
had one million dollars to help support the chair. One year, in early
December, Ron attended a meeting with me in my office and handed me an
envelope with roughly \$100,000 worth of America
On Line shares of stock. I thought for a moment, there's a Delta jet
with connecting flights to Hawaii leaving in about an hour. I could
promptly turn the envelope over to the university's chief fundraiser at
the time, Pat Martin, who was also attending the meeting. Or I could
excuse myself, take the envelope with me ostensibly to return in a
moment with the shares in a more carefully labeled envelope, and
instead catch that jet.$\ldots$ Later that morning I noticed the sky
was blue and clear as the engines roared and we took off to the west.

\textbf{Samaniego:} Myles, that could be the beginning of the next
great American novel!

\section*{Editorial Activities}

\textbf{Samaniego:} You served as the Editor of the Theory and Methods
Section of JASA in 1994--1996. I know this is an extremely labor-intensive
job. You seemed to thrive on the experience. What did you enjoy most
about it?

\textbf{Hollander:} I had a great board of associate editors,
including you, and you gave me the luxury of three reviews per paper.
I~liked working with the board. I~also enjoyed reading the
submissions---one year I had 503!---and the reviews. I tried to
encourage authors, and with the reviews, improve the papers. Even if a
paper was declined, I wanted the disappointed author to feel his/her
paper was treated with respect and got a fair shake. I helped to get a
page increase and in some of my issues I had over 30 papers in
Theory and Methods. I also increased the T\&M acceptance rate to around
30\%. I suspect it is significantly lower now. It was just a great
experience. Many nights and weekends I would bring a stack of folders
home. If an AE was very tardy, I threatened to send in a SWAT team or
toss him in a dark cellar until I received the reviews. One of my main
goals was to make the papers readable and understandable. I insisted
the authors write for the readers. I believe that was a mark of my
editorship and your editorship, Frank, as well.

I enjoyed being JASA editor and a JASA AE before the
editorship. I served on the boards of Paul Switzer, Ray Carroll and Ed
Wegman, learned a lot from them, and was grateful for the
opportunities. I've also continued with editorial activities after my
JASA term ended. In 1993, the first volume of the \textit{Journal of
Nonparametric Statistics}, founded by Ibrahim Ahmad, appeared and I
have been a board member since then, with one break. In 1995, Mei-Ling
Lee launched \textit{Lifetime Data Analysis}, and I have been a board
member since the beginning. Both of those journals publish important
papers and the profession should be, and I believe is, grateful to
Ibrahim and Mei-Ling for their visions and dedicated work.

\section*{The Future}

\textbf{Samaniego:} In 2003, you received the Noether Senior Scholar
Award for your work in nonparametric statistics. That must have been
extremely satisfying. What do you see as the important open problems
that current and future researchers in this area might wish to focus
on?

\textbf{Hollander:} The Noether Award is very special to me. The list
of awardees consists of distinguished people with major accomplishments
in nonparametrics and I am very grateful for the honor. The awardees
thus far are Erich Lehmann, Bob Hogg, Pranab Sen, me, Tom
Hettmansperger, Manny Parzen, Brad Efron and Peter Hall.

Stephen Hawking, the great physicist, says you cannot predict
the great innovations in the future; that's partially why they are
termed great innovations. If, however, Dennis Lindley is correct about
this being a Bayesian century, and it seems to be going in that
direction, then I would like nonparametrics to play a major role. Thus
I would wish for new, important innovations in Bayesian nonparametrics.
In my department we have at least three faculty members, Anuj
Srivastava, Victor Patrangenaru and Wei Wu, working in image analysis,
target recognition, face recognition and related areas. I~would like to
see nonparametric developments in these areas which are obviously
important in many arenas including medical diagnoses and national
security.

As a field, I'm glad we are pushing hard in interdisciplinary
work, and it's good for our future role in science. It's valuable for
the quality of research in the outside areas with which we participate
and for scientific research overall. I'm hopeful statisticians will
contribute significantly to many of the important open questions in
other fields and many already do. In academic settings, it's critical
that university administrations recognize the importance of strong
statistical support raising the quality of research.

I want to be surprised in the future but, like Hawking says,
it's hard to guess at the surprises. What do you think, Frank?

\textbf{Samaniego:} In the 20th century, especially from say, 1940 to
1990, the mathematical aspects of statistics were emphasized in both
teaching and research. Mathematical statistics was prime. The power of
computation changed that considerably. Then, applied problems, real
applications with large and complex data sets, changed it even more.
Today, there are areas like data mining that are of great interest and
importance but haven't yet been mathematized. I wonder if it's just too
early to mathematize challenging problems like these. I'm guessing that
some sort of theory of optimality, some sense of what's good and what's
better than something else, will be part of the future development of
these evolving problem areas. It's just simply too hard to do this with
tools we have available now.

\textbf{Hollander:} It is true that you can do a lot of things now
with computer-intensive methods and not worry about getting the exact
results. It's a little reminiscent of when Karl Pearson was classifying
curves. There are a lot of data-based methods, but the mathematical
foundations may have to be solidified. I think now that we are pushing
applied stuff, computer-intensive methods, we can get results
relatively easily, for example, nonparametrics with bootstrapping and
Bayesian methods with MCMC. We may have to go back a little bit and
shore up some of the methods, study their performance and properties as
you suggest. But I think that will be considered only by theoretical
statisticians. The computer-intensive surge is of course going to keep
rolling, yield many new discoveries, and is great for the field.

\section*{Other Interests}

\textbf{Samaniego:} You have many collateral interests, not the least
of which is baseball. You once told me that you were as pleased with
your published letters to the Editor of \textit{Sports Illustrated} as
you were with many of your professional accomplishments. Tell me about
your interest in the Dodgers and in sports in general.

\begin{figure*}

\includegraphics{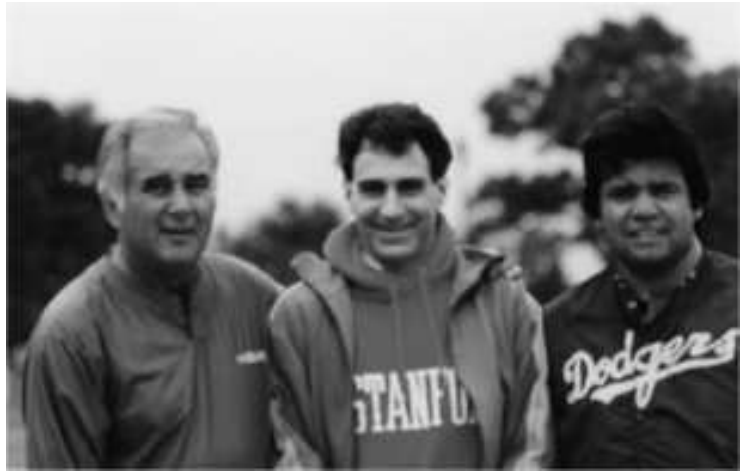}

\caption{Myles Hollander flanked by Dodger pitching coach Ron
Perranoski and Dodger pitcher Fernando Valenzuela, Dodgertown, Vero
Beach, Florida, 1986.}
\end{figure*}

\textbf{Hollander:} I was just kidding about the importance of the
SI letters. Getting a statistical paper published is much more
satisfying and represents a long-term and dedicated effort. However,
the letters arose this way. My friend Bob Olds, a psychiatrist in St.
Augustine, used to live in Tallahassee and write columns for the local
newspaper. His future wife, Ann, took a few classes from me when she was
an undergraduate at FSU. Bob sent a few letters to SI and they were not
published. He is a wonderful writer, much better than me, but just for
fun I submitted two and, surprisingly, both were
accepted. The first was about Dodger pitcher Fernando Valenzuela during
a period of Fernandomania in LA. The second was a comparison of the
Stanford and Florida State marching bands. The latter was prompted by
that bizarre play in November, 1982, at the Cal--Stanford Big Game. You
may recall that the Stanford band prematurely went on the field near
the end of
the game thinking Stanford had won and they inadvertently ended up as
blockers on Cal's game-ending touchdown.

My interest in the Dodgers came about naturally during my
childhood in Brooklyn. During my summers in high school most of my
friends were away at what was then called sleep-away camp. My parents
could have afforded to send me, and I wanted to go, but I was an only
child and they liked having me around. So I had summer jobs in the city
and then on weekends, and on some evenings, I would walk to Ebbets
Field, sit in the bleachers or the grandstand, and watch the Bums, as
they were affectionately called. This was the era of Jackie Robinson
who displayed tremendous courage when he broke the color line in
baseball. Branch Rickey, the Dodgers' General Manager at the time, also
deserves a lot of credit for giving Robinson the opportunity. I enjoyed
talking baseball to strangers at the game, seeing Afro-Americans and
Caucasians get along, and I loved the teamwork on the field. I've lived
my life with respect for people from all walks of life, from different
backgrounds and cultures, and the Dodgers played a role in teaching me
that. During my years as chair, I tried
to instill the same kind of teamwork in the department.

I liked playing sports, mostly basketball, baseball and
tennis. In my childhood, on the streets of Brooklyn, I played city
sports like punchball and stickball. I also played basketball in
schoolyards and baseball at the Parade Ground in Brooklyn. I played
some tennis in high school but didn't get reasonably skilled at it
until the early '70s.

\textbf{Samaniego:} One of the things that I've noticed about you over
the years is that you and Glee like to go down to Vero Beach to see
some spring training games. How long has that tradition been going on?

\textbf{Hollander:} I would say it dates back to the '70s, almost the
time we first came to Tallahassee. We came to Tallahassee in 1965. We
used to go to see the Dodgers. It was a different era. We could
actually go up to them and talk
to them and chat about baseball, whereas today they're much more
isolated. There are fences. I had some good conversations with players
over the years. I remember once we went to Vero Beach and the game was
rained out. It was a game against Boston. Fernando Valenzuela was
practicing with his pitching coach, Ron Perranoski. They were tossing
the ball on a practice
field so Glee and I went up to them and started talking to them and
they also posed for pictures. We have many pictures from those years.
One with our sons Layne and Bart and Hall-of-Fame Dodger pitcher Sandy
Koufax is here on the office wall.

\textbf{Samaniego:} Has any of your work involved sports in
statistics?

\textbf{Hollander:} I haven't done serious sports statistics like the
type that interests the sports statistics section of the ASA. In the
early '70s, however, Woody Woodward, who had been a player on FSU's
baseball team, came to my office for help on the design and analysis of
a study on different methods of rounding first base. I helped him and
it became part of his master's thesis. Later Doug and I put the example
in our nonparametrics book. In appreciation for the consulting,
Woodward sent me a baseball glove from spring training when he was a
member of the Cincinnati Reds. I used it when I played intramural and
city league softball at FSU and I still take it with me to spring
training games and major league games, hoping to catch a foul ball.

\begin{figure}

\includegraphics{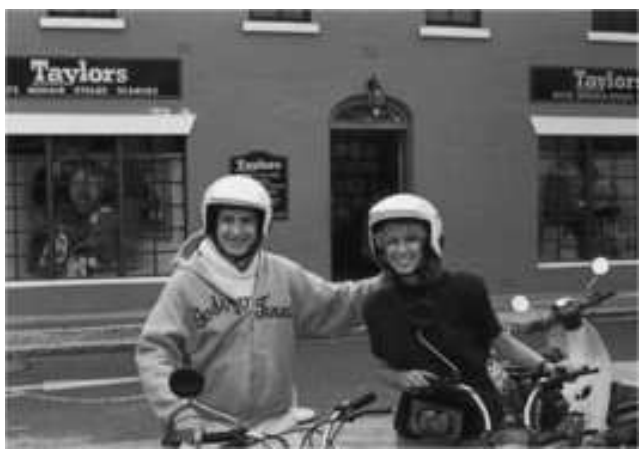}

\caption{Myles and Glee Hollander riding mopeds in Bermuda, 1994.}
\end{figure}

\begin{figure}

\includegraphics{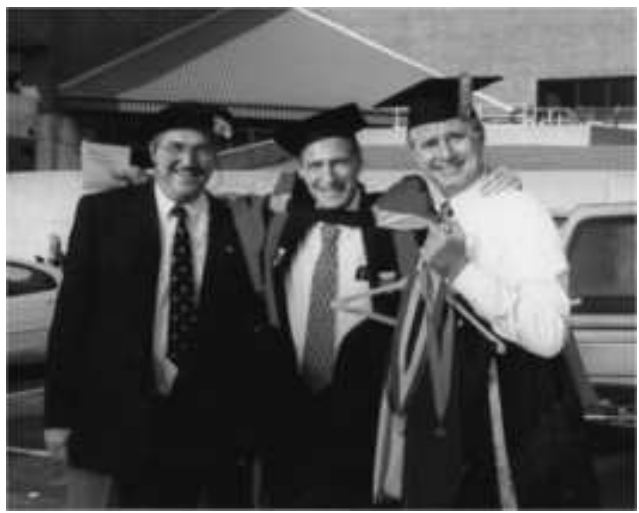}

\caption{Mathematics chair DeWitt Sumners, Dean Donald Foss
and Myles Hollander on the occasion of Myles' Robert O. Lawton
Distinguished Professor Award, Tallahassee, 1998.}
\end{figure}

\begin{figure}

\includegraphics{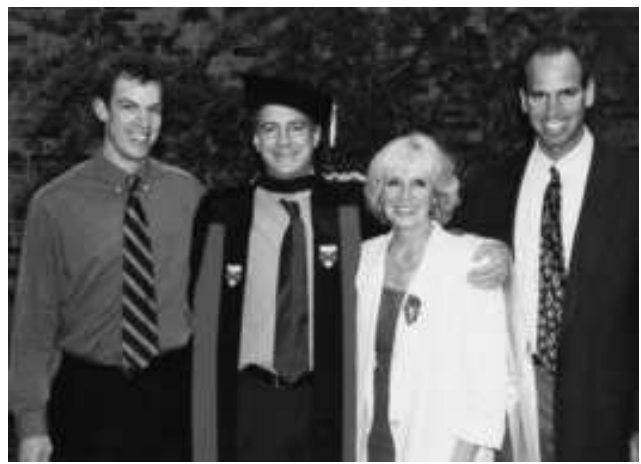}

\caption{Myles and Glee Hollander with their sons Layne Q and
Bart Q after the Lawton Award Luncheon, Tallahassee, 1998.}
\end{figure}

\begin{figure*}

\includegraphics{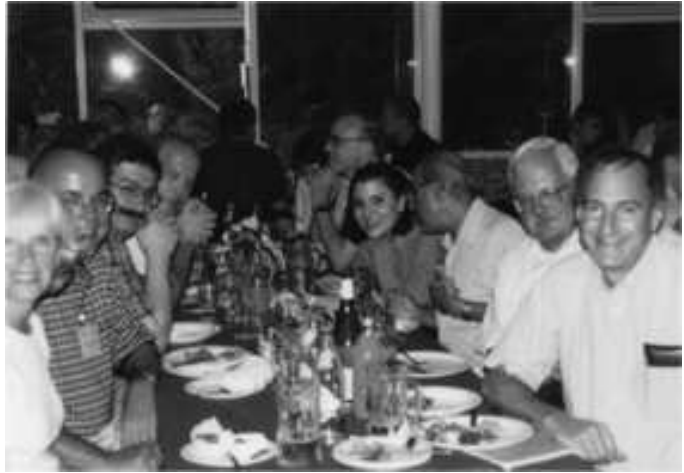}

\caption{Glee Hollander, Myles Hollander, Frank Samaniego,
Henry Block, Nozer Singpurwalla, Refik Soyer, Elena Samaniego at the
10th INFORMS Applied Probability Conference, University of Ulm, Ulm,
Germany, 1999.}
\end{figure*}

\textbf{Samaniego:} I'm visiting FSU on the occasion of a conference
honoring your contributions to statistics and your department and
university and commemorating your upcoming retirement. I know that
you're looking forward to spending more time with family. I'm sure your
sons and your grandkids will soak up plenty of your freed-up time. Any
special plans?

\textbf{Hollander:} You're right. Glee and I do want to spend more
time with our sons Layne, and his children Taylor and Connor, and Bart,
his wife Catherine, and their children Andrew, Robert and Caroline.
One set lives in Plantation, Florida, one in Amherst, Massachusetts.
That will prompt some traveling. Also, Glee has siblings in Hilton
Head, South Carolina and Spokane,
Washington and I have family in LA, so we will get around. I also hope
to go to a few statistical events. I~love the international travel to
conferences. You and I often attend the ones featuring reliability with
the usual reliability
club, Ingram Olkin, Nozer Sinpurwalla, Allan Sampson, Nancy Fluornoy,
Henry Block, Edsel Pe\~na, Mark Brown, Phil Boland, Jim Lynch, Joe
Glaz, Nikolaos Limnios, Misha Nikulin, many more.

Glee and I own a beach house at Alligator Point, Florida.
It's about an hour drive from our home in Tallahassee. We expect to be
there a lot, walk on the beach, take bike rides to the western end of
the point where there is a bird sanctuary, read novels, and so forth.

\textbf{Samaniego:} I've got to believe that you have at least one
more book in you. Do you hope to do some writing once you are
officially retired?

\textbf{Hollander:} Possibly I'll write a book. Realistically, I think
it's more likely I'll stay involved by writing a paper every now and
then and recycling back to FSU from time to time to teach. Lincoln
Moses said, ``There are no facts for the future.'' Despite being a
statistician, I can't predict.

\textbf{Samaniego:} You've had a long and productive career as a
research statistician. Looking back, what would you say is your
``signature'' result?

\textbf{Hollander:} I'll interpret the word ``signature'' literally
and take the opportunity to say I greatly enjoyed the work we did
together on your elegant concept of signatures in reliability theory
during my sabbatical visit to UC Davis in Spring, 2006 (Hollander and
Samaniego, \citeyear{hs07}). For comparison of two coherent systems, each having
i.i.d. components with a
common distribution $F$, we suggested the distribution-free measure
$P(X < Y)$ where $X$ is the lifelength of system 1 and $Y$ is the
lifelength of system 2. We found a neat way to calculate the measure
directly in terms of the systems' signatures and probabilities
involving order statistics. Among other things, we resolved the
noncomparability issues using stochastic ordering, hazard rate
ordering and likelihood ratio ordering that you (Kochar, Murkerjee
and Samaniego, \citeyear{kms99}) encountered for certain pairs of systems.

In the bigger picture, my signature career quest was to
promote nonparametric statistics, bring it into other areas, get more
people to use it, and get students to study the subject and make
contributions to the field.

\begin{figure}

\includegraphics{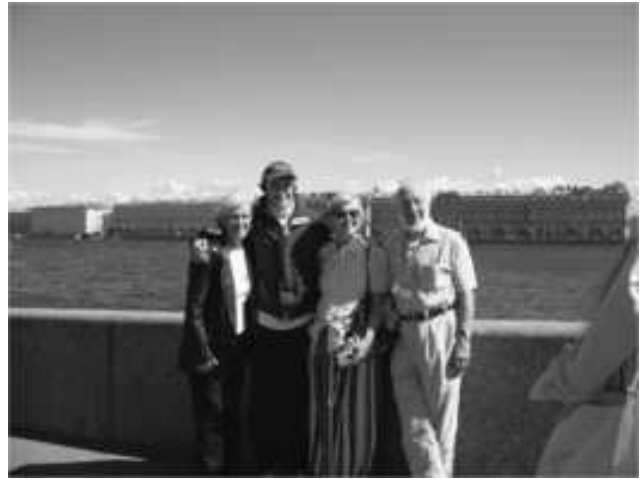}

\caption{Myles and Glee Hollander with Ingram Olkin and Nancy
Fluornoy across from the Winter Palace while attending the 5th St.
Petersburg Workshop on Simulation, St. Petersburg, Russia, 2005.}
\end{figure}

\begin{figure*}

\includegraphics{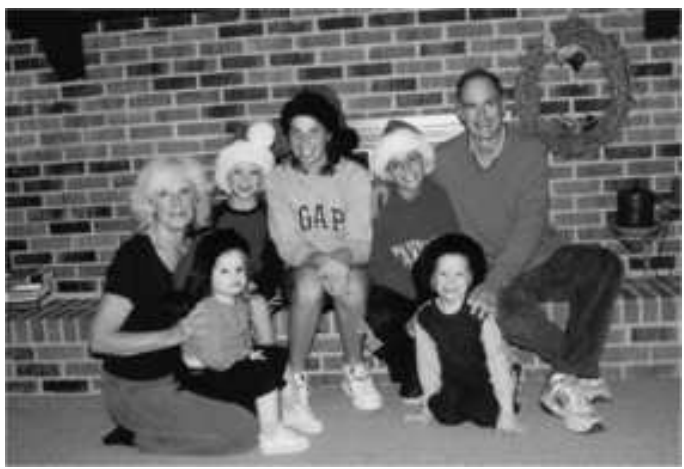}

\caption{Myles and Glee Hollander with their grandchildren
Caroline, Andrew, Taylor, Connor and Robert, Tallahassee, 2006.}
\end{figure*}

\textbf{Samaniego:} It seems that, over the period of your career,
nonparametric methods have become more and more important and
pervasive. There is no question that your work has helped that
direction significantly.

\textbf{Hollander:} Thank you, Frank. When I look in journals there are
a lot of papers that are nonparametric in nature and the adjective
nonparametric does not appear in the titles. It's just a natural way to
start a problem now, letting the underlying distributions be
arbitrary.

\textbf{Samaniego:} You've worked with some of the legendary figures
in our discipline including Ralph Brad\-ley, Frank Wilcoxon, Richard
Savage. These colleagues, and others, have played important roles in
your professional evolution. How did the general environment at Florida
State help shape your career?

\textbf{Hollander:} I came to Florida State because of Ralph, Frank
and Richard. They were three luminaries in nonparametric statistics and
I wanted to do nonparametrics research. Frank and I shared an office;
he and his wife Feredericka and Glee and I became friends, but he died
three months after I arrived. I never did research with Frank, Ralph
or Richard. But I was close to them. Ralph and his wife Marion and
Richard and his wife Jo Ann were always friendly to Glee and me.
Although I didn't write with Richard, up to the time he left for Yale
in 1973 he carefully read each one of my technical reports and often
made valuable suggestions. I gave the Bradley lectures at the
University of Georgia in 1999, and after Ralph passed away, I was asked
by his family to deliver a eulogy at his memorial service in Athens
which I did, with pleasure.

The environment at FSU was dedicated to research and I liked
that. I came to a place where that was the top priority. Also I came in
1965, only six years after the department was founded by Ralph, so
there was the excitement of building. As it turned out, I was there
when the first Ph.D. graduated and thus far I have seen all of our
Ph.D. students graduate.

\textbf{Samaniego:} My recollection is that there's a famous quote
attributed to you about the discipline of statistics. Tell me about
it.

\textbf{Hollander:} The saying is: ``Statistics means never having to
say you're certain.'' I saw the movie ``Love Story'' in 1971. It was a
big hit. It was based on a book of the same title by Erich Segal. I
read the book after I saw the movie. As the title indicates, it's a
love story. A wealthy Harvard law student, Oliver Barrett, falls for a
poor Radcliffe girl, Jennifer Cavilleri, and eventually they marry. At
one point, after a spat, Oliver apologizes and Jenny replies, ``Love
means never having to say you're sorry.''

In statistics we give Type-I and Type-II error probabilities,
confidence coefficients, confidence bands, false discovery rates,
posterior probabilities and so forth, but we hedge our bets. We assess
the uncertainty. With the movie fresh in my mind I transformed Segal's
phrase to ``Statistics means never having to say you're certain.''

\textbf{Samaniego:} Thanks, Myles. This excursion has been most
enjoyable!

\textbf{Hollander:} Frank, we have had a long friendship that has
stood the test of a continental divide between us. I look forward to
its future pleasures. Thank you for the conversation. It was highly
enjoyable and I'm grateful for the opportunity to interact in this
manner and offer my musings.

\section*{Acknowledgments}

Myles Hollander and Frank Samaniego thank Pamela McGhee, Candace Ooten
and Jennifer Rivera for carefully transcribing the conversation.

\end{document}